\documentclass[5p,final]{elsarticle}
\usepackage{xurl}
\usepackage{amsfonts}
\usepackage{amssymb}
\usepackage{amsmath}  
\usepackage{amsthm}
\usepackage{enumitem} 
\usepackage{geometry}
\usepackage{graphicx}
\usepackage{xfrac}
\usepackage[colorlinks=true, allcolors=blue]{hyperref}
\usepackage[utf8]{inputenc} 
\usepackage[english]{babel} 
\usepackage{url}
\usepackage{wrapfig}
\usepackage{subfiles}
\usepackage{multirow}
\usepackage{cleveref}
\usepackage{balance}
\usepackage{svg}
\usepackage{booktabs}
\usepackage{tikz}
\usetikzlibrary{shapes.geometric, arrows}
\usepackage{tikz-3dplot}
\usepackage[squaren]{SIunits}
\usepackage{caption}
\usepackage[normalem]{ulem}
\usepackage{subcaption}
\usepackage{lipsum}
\usepackage{todonotes}
\usepackage{float}

\newcommand{\bc}{\begin{center}}
\newcommand{\ec}{\end{center}}
\newcommand{\be}{\begin{equation}}
\newcommand{\ee}{\end{equation}}
\newcommand{\bea}{\begin{eqnarray}}
\newcommand{\eea}{\end{eqnarray}}
\newcommand{\beq}{\begin{eqnarray*}}
\newcommand{\eeq}{\end{eqnarray*}}
\newcommand{\bv}{\left( \begin{array}{c} }
\newcommand{\ev}{\end{array} \right) }

\newcommand{\Var}{\mbox{Var}}
\newcommand{\Cov}{\mathbb{C}\mathrm{ov}}

\usepackage{natbib}
\begin{document}

\title{Correlation emergence and the Epps effect in two coupled limit order books}

\author[unsw-mathstats]{Chris Angstmann}\ead{c.angstmann@unsw.edu.au}
\address[unsw-mathstats]{School of Mathematics and Statistics, University of New South Wales, Sydney, NSW 2052, Australia}
\author[uct-sta]{Tim Gebbie} \ead{tim.gebbie@uct.ac.za}
\address[uct-sta]{Department of Statistical Sciences, University of Cape Town, Rondebosch 7701, Western Cape, South Africa}

\begin{abstract}
We give a unified analytic account of correlation emergence and the Epps effect in two coupled limit order books. The Epps effect is the empirical reduction in measured cross-asset log-return correlation at short aggregation scales, or equivalently the recovery of measured correlation as the aggregation interval increases. The model starts from a fixed-grid discrete random walk for order flow in operational time, with creation, cancellation and diffusion. A pair-trader coupling between the books is introduced at the level of order creation, and calendar time is imposed through separate observation clocks. We clarify how the operational-time model reduces to coupled reaction--diffusion equations with a moving reaction boundary defining the model log-mid-price. Using a regularised local-response representation of the coupling, we derive approximate closed-form expressions for realised correlations as a function of aggregation time. Here the Epps effect is shown to arise from two distinct mechanisms: asynchronous observation clocks (subordination), finite coupling response times, and their combination.
\end{abstract}

\begin{keyword}
Limit order book \sep Reaction--diffusion \sep Epps effect \sep Subordination \sep Market microstructure
\end{keyword}

\maketitle

\section{Introduction} \label{sec:introduction}

Financial markets operate through intricately coupled collections of order books. These can be distinguished as being either lit or latent order books where both can contribute to the dynamics of price emergence. Lit order books contain displayed orders that market participants can observe and trade against. Their transparency supports liquidity by making available orders and quoted prices visible. Lit order books match displayed orders according to rules such as price-time priority \citep{ohara1995}. Hidden or non-displayed orders are orders that have already been submitted to a market, but are not displayed in full. 

The latent order book \citep{toth2011} used here is slightly different and something more fundamental. It represents potential supply and demand from trading intentions that have not yet been expressed as orders but are latent in the broader market. \citet{donier2015} introduced the reaction--diffusion representation of the latent order book. \citet{Gant2022b} implemented a stochastic finite-difference solution \citep{angstmann2019advection,angstmann2016} and calibrated the resulting market model. \citet{DianaGebbie2025} then developed the single-book DTRW construction with anomalous diffusion, non-uniform event times and simulated price impact. \citet{BauerDianaGebbie2024} then coupled two simulated lit order books through a pair-trader mechanism and showed numerically that correlation and an Epps effect emerge from the coupled asynchronous dynamics. 

The Epps effect is the empirical decline in measured cross-asset return correlation as the sampling or aggregation interval is shortened, first documented by \citet{epps1979}. The Epps effect has been observed in various empirical studies of financial markets \citep{toth2009,mastromatteo2011}. Here we give an analytical reaction-front treatment of the clock and coupling-response mechanisms, together with their combination to demonstrate the Epps effect. This is done using a clock separation construction where we first construct the reaction boundary in operational time and then subordinate it to some calendar time.

Correlation can then be shown to emerge from the order-flow coupling between the books. The coupling is phenomenological and represents trader interactions; it is not introduced as a prespecified dependence between observed event types or price processes. In a Hawkes model of order-book events, scalar log-return correlation appears after events are mapped into price changes, while the cross-dependence is specified through the event taxonomy and selected kernels \citep{bacry2015,chang2025discrete}. Our construction places the cross-book mechanism directly with a mesoscopic pair-trader coupling whose price effect passes through the order-density fields and moving reaction fronts.

Because the analytical treatment separates the reaction-boundary dynamics in operational time from the asynchronous clocks through which prices are observed, this identifies the attenuation due to clock overlap separately from the finite response of the pair-trader coupling. This leads naturally to a subordinated limiting model. The numerical correspondence is stated at the reaction-front response level in \ref{app:numerical-reaction-front-correspondence}.

The numerical study tests the clock and coupling mechanisms separately before combining them without refitting. The implementation evolves the order books on a uniform operational grid, applies the directed coupling through the current reaction-front translation modes, and observes the completed boundary paths through book-specific previous-refresh clocks. Section~\ref{sec:numerical-implementation} gives the numerical comparisons. 

%The operational and calendar-observation algorithms are supplied in the standalone supplementary algorithms document in the paper reproducibility code release \cite{AngstmannGebbie2026ReproducibilityCode}.

\section{The coupled limit order books} \label{sec:2}

We separate the order-book dynamics from the clocks through which they are observed. Let
\begin{equation}
    x_i=i\Delta x,
    \qquad
    u_n=n\Delta u,
    \qquad n\in\mathbb N,
\end{equation}
where $\Delta x>0$ is the fixed log-price-grid spacing, $\Delta u>0$ is the fixed operational-time step, $u$ is operational time and $t$ is reserved for calendar time. At each operational-time step, orders may be created or cancelled, and existing orders may shift by one log-price level. For a single book, the microscopic dynamics is a discrete-time random walk in log-price space with killing and source terms \citep{angstmann2016,angstmann2019advection,DianaGebbie2025}.

Let $\varphi_n^{i(j)}$ denote the signed order density at grid point $x_i$ for asset $j$ at operational-time index $n$, and set
\begin{equation}
    0\leq r\leq1,
    \qquad
    |F_n^{(j)}|\leq r.
\end{equation}
For the continuum construction take $r>0$. The incoming nearest-neighbour weights are $\tfrac12(r+F_n^{(j)})$ and $\tfrac12(r-F_n^{(j)})$, where $F_n^{(j)}$ is the signed stochastic jump bias. The quantity $K_{n-m}^{(\alpha_u)}$ is the Sibuya memory kernel of order $0<\alpha_u\leq1$ \cite{angstmann2016}, and $\nu_j$ is the cancellation rate in book $j$. The exponent $\alpha_u$ belongs to the order-density dynamics in operational time; the calendar-time clock exponent $\alpha_t$ is introduced separately below.

The total source increment is then:
\begin{equation}
    c_n^{i(j)}
    =s_n^{i(j)}+\delta_n^{i(j)}
    +\sum_{k\ne j}\ell_{\varepsilon,n}^{i(j,k)},
    \label{eq:discrete-total-source}
\end{equation}
where $s^{(j)}$ is the one-book order source, $\delta^{(j)}$ is an external shock and $\ell_\varepsilon^{(j,k)}$ is the ordered-pair source acting on book $j$ in response to book $k$, with regularisation parameter $\varepsilon$. On the fixed operational-time and log-price lattice
\begin{equation}
\begin{aligned}
\varphi^{i(j)}_n
&= \sum_{m=0}^{n-1} K_{n-m}^{(\alpha_u)}
 e^{-\nu_j (u_{n-1}-u_m)} \bigg[
\tfrac{1}{2}\big(r + F^{(j)}_{n-1}\big)\varphi^{i-1(j)}_{m} \\
&\quad + \tfrac{1}{2}\big(r - F^{(j)}_{n-1}\big)\varphi^{i+1(j)}_{m}
- r \varphi^{i(j)}_{m}
\bigg] \\
&\quad + e^{-\nu_j\Delta u}\varphi^{i(j)}_{n-1}
+ c_{n-1}^{i(j)}\Delta u.
\end{aligned}
\label{eq:operational-DTRW-update}
\end{equation}
The common-cancellation-rate case is recovered by setting $\nu_1=\nu_2=\nu$. A non-zero jump bias surviving the continuum scaling would generate an advection term. The density equation used below takes the centred limit $F_n^{(j)}=0$.

\subsection{Coupled source terms} \label{ssec:coupledPDE}

An ordered-pair term $\ell_\varepsilon^{(j,k)}$ changes book $j$ in response to book $k$; single-book quantities retain only the index $j$. Boundary conditions ensuring conservation of the integrated signed density make it convenient to consider lit order-book sources \citep{Gant2022b,DianaGebbie2025,BauerDianaGebbie2024}:
\begin{equation}\label{eq:source-term}
s^{(j)}(x,u) = - \lambda_j \mu_j (x - p^{(j)}(u)) e^{-\mu_j\left(x - p^{(j)}(u)\right)^2}.
\end{equation}
Here $p^{(j)}(u)$ is the reaction-boundary log-price defined below, $\lambda_j$ is the source-strength parameter, and $\mu_j$ is the source-shape parameter. The coupling represents an external pairs trader who observes the two books and places side-selective orders in response to their directed log-mid-price difference. We focus on two books, although the ordered-pair notation extends directly to more assets.

\subsection{Moving reaction boundary and calendar clock}

The operational-time model log-mid-price for book $j$ is the unique reaction-boundary zero crossing
\begin{equation}
    p^{(j)}(u)=\{x:\varphi^{(j)}(x,u)=0\}.
    \label{eq:operational-price}
\end{equation}
We assume constant annihilation rates,
\begin{equation}
    a^{(j)}(x,u)=\nu_j,
\end{equation}
and write the total source as
\begin{equation}
    c^{(j)}(x,u)=s^{(j)}(x,u)+\delta^{(j)}(x,u)
    +\sum_{k\ne j}\ell_\varepsilon^{(j,k)}(x,u).
\end{equation}
Order-book shocks $\delta^{(j)}$ can be used to estimate price impact \citep{DianaGebbie2025} by introducing a shock of size $Q$ and measuring the resulting log-price change. Both lit and latent order-book sources contribute to price emergence: lit sources have vanishing boundary conditions, whereas latent sources may have finite boundary values.

Let $T_j(u)$ be a non-decreasing process mapping operational time into calendar time, and define its inverse clock by
\begin{equation}
    E_j(t)=\inf\{u\geq0:T_j(u)>t\}.
    \label{eq:inverse-clock}
\end{equation}
The calendar-time order density and log-price are defined pathwise by
\begin{equation}
    \Phi^{(j)}(x,t)=\varphi^{(j)}\!\left(x,E_j(t)\right),
    \qquad
    P^{(j)}(t)=p^{(j)}\!\left(E_j(t)\right).
    \label{eq:field-price-subordination}
\end{equation}
Under uniqueness of the zero crossing, $P^{(j)}(t)=\{x:\Phi^{(j)}(x,t)=0\}$. Thus $p^{(j)}(u)$ and $P^{(j)}(t)$ are log-prices; exponentiating them gives the corresponding model price levels. The common-clock case is $E_1=E_2$. Asset-specific clocks allow $E_1\ne E_2$ and are required for a separate clock-overlap attenuation mechanism.

\section{From discrete order flow to a continuum limit}
The fixed-grid operational-time limit is taken under
\begin{equation}\label{eq:d}
    D_{\alpha_u}
    =\lim_{\substack{\Delta x\to0\\\Delta u\to0}}
    \frac{r}{2}\frac{\Delta x^2}{(\Delta u)^{\alpha_u}},
\end{equation}
where $D_{\alpha_u}$ is the diffusion constant\footnote{With dimensions $[D_{\alpha_u}]=[\mathrm{log\text{-}price}]^2[\mathrm{operational\ time}]^{-\alpha_u}$}. Following \citet{angstmann2016,angstmann2019advection}, let $\varphi_\Delta^{(j)}(x,u)$ be the lattice-dependent approximating function satisfying $\varphi_\Delta^{(j)}(x_i,u_n)=\varphi_n^{i(j)}$. Define the operational-time cancellation survival function $\theta_j(u)=e^{-\nu_ju}$. The continuum order density then satisfies
\begin{equation}
\begin{aligned}
\partial_u \varphi^{(j)}
&= D_{\alpha_u}\partial_x^2\left[
\theta_j(u)\,{}^{\mathrm{RL}}D_u^{1-\alpha_u}
\left(\frac{\varphi^{(j)}}{\theta_j(u)}\right)
\right]\\
&\quad-\nu_j\varphi^{(j)}+c^{(j)}(x,u).
\end{aligned}
\label{eq:operational-continuum-density}
\end{equation}
The survival weighting is required when $0<\alpha_u<1$ and $\nu_j>0$. For $\alpha_u=1$, the memory operator has order zero and Equation~\eqref{eq:operational-continuum-density} reduces to the ordinary operational-time reaction--diffusion equation. For $\nu_j=0$, it reduces to the unweighted fractional diffusion form.

The reaction-boundary log-price is
\begin{equation}
    p^{(j)}(u)=\{x:\varphi^{(j)}(x,u)=0\}.
\end{equation}
This moving boundary separates buy- and sell-dominated regions of the order book. The one-front construction does not resolve the best bid and ask separately. The zero crossing is the model log-mid-price; exponentiating it gives the corresponding model price level, which at this reduced level is also used to approximate the realised transaction price. Its calendar-time projection is obtained from Equation~\eqref{eq:field-price-subordination}.

When $E_j$ is an inverse-stable clock of order $0<\alpha_t\leq1$, a closed calendar-time fractional equation follows directly for a linear operational-time generator. With state-dependent or nonlinear coupling sources, the subordinated field remains well defined by Equation~\eqref{eq:field-price-subordination}, but a closed fractional PDE requires a separate linearisation, weak-limit or closure assumption. The exponent $\alpha_t$ is therefore not inferred from $\alpha_u$. The identification $\alpha_u=\alpha_t$ is thus an optional modelling restriction.

\subsection{Regularised analytic coupling notation}
\label{ssec:regularised-analytic-coupling}

For the analytical derivation we use a bounded regularised representation of the directed pair-trader mechanism. The numerical implementation instead applies the ordered response rate directly through the current receiving-front translation mode. These source fields are not pointwise identical; their reduced reaction-front correspondence is stated in \ref{app:numerical-reaction-front-correspondence} and~\ref{app:first-moment-correspondence}.

Write the local coordinate and directed spread as
\begin{equation}
\begin{aligned}
    y_j(x,u)&:=x-p^{(j)}(u),\\
    z_{jk}(u)&:=p^{(j)}(u)-p^{(k)}(u),
    \qquad z_{kj}(u)=-z_{jk}(u).
\end{aligned}
\end{equation}
The analytic source-shape kernel is written
\begin{equation}
    q_j(y)=-\lambda_j\mu_j ye^{-\mu_j y^2},
    \label{eq:qj-analytic}
\end{equation}
where $\lambda_j>0$ and $\mu_j>0$. The first-moment identity requires $q_j$ to be odd, integrable and to have a finite first moment.

The regularised pair-trader coupling from book $k$ into book $j$ is
\begin{equation}
    \ell_\varepsilon^{(j,k)}(x,u)
    =\gamma_{jk}z_{jk}(u)q_j\!\left(y_j(x,u)\right)
      W\!\left(y_j(x,u),z_{jk}(u);\varepsilon\right),
    \label{eq:regularised-coupling}
\end{equation}
where $\gamma_{jk}\geq0$ is a coupling-strength parameter and $W$ is a bounded side-selection function. A smooth choice \cite{OlssonKreiss2005,XiaoHonmaKono2005} is
\begin{equation}
    W(y,z;\varepsilon)
    =\frac{1}{2}\left[1+\tanh\!\left(\frac{yz}{\varepsilon}\right)\right],
    \qquad \varepsilon>0.
    \label{eq:side-selection-function}
\end{equation}
Since $y$ and $z$ both have units of log-price, $\varepsilon$ has units of log-price squared. For fixed $z\ne0$, the transition width in $y$ is $w_\varepsilon(z)\sim\varepsilon/|z|$. We use the reference-scale parameterisation
\begin{equation}
    \varepsilon=|z|_{\mathrm{ref}}w_{\mathrm{ref}},
    \qquad
    \Delta x\ll w_{\mathrm{ref}}\ll L_{q,j},
    \qquad
    L_{q,j}\sim\mu_j^{-1/2}.
\end{equation}
Thus $z_{jk}>0$ selects the $y_j>0$ side of book $j$, while $z_{jk}<0$ selects the $y_j<0$ side. The corresponding hard-threshold source uses $W_0(y,z)=\mathbf{1}_{\{yz>0\}}$.

The smooth and hard-side continuum sources have the same raw first spatial moment, as shown in \ref{app:first-moment-correspondence}. Conversion of that moment into the ordered reaction-front rate uses the normalised closure and validity assumptions in \ref{app:coupling-only}; the numerical implementation then realises the declared rate through the receiving-front translation mode.

\section{Analytic decomposition of the Epps effect}

The operational-time log-prices $p^{(j)}(u)$ and their calendar-time images $P^{(j)}(t)$ are from Equation~\eqref{eq:field-price-subordination}. Now, let $R^{(j)}_{\Delta t}(t)=P^{(j)}(t+\Delta t)-P^{(j)}(t)$ denote the calendar-time log return over aggregation interval $\Delta t$.
Then the realised correlation \cite{BarndorffNielsenShephard2004RealizedCovariation} is
\begin{equation}
 \rho_{\Delta t}=\frac{\Cov(R^{(1)}_{\Delta t},R^{(2)}_{\Delta t})}{
 \sqrt{\Var(R^{(1)}_{\Delta t})\Var(R^{(2)}_{\Delta t})}}. \label{eq:realised-corr}
\end{equation}

Equation \eqref{eq:realised-corr} is the population analogue of the realised-correlation estimator used in the high-frequency realised-covariation literature \citep{BarndorffNielsenShephard2004RealizedCovariation}. When the two assets are observed on non-synchronous event clocks, the relevant overlap structure is closely related to non-synchronous covariance estimation \citep{hayashi2005}.

We write $\rho_u$ for the instantaneous correlation of the operational-time innovations and reserve $\rho_\infty$ for the large-aggregation correlation level of the observed log-price model. These quantities coincide only in the clock-only benchmark after the overlap factor tends to one.

\subsection{Subordination-only mechanism}

If log-prices evolve in operational time but are observed through independent clocks, the correlation is reduced
by limited clock overlap.
One finds
\begin{equation}
 \rho_{\Delta t}^{\mathrm{sub}}=\rho_u \, \mathcal{A}_{12}(\Delta t),
\end{equation}
where $\mathcal{A}_{12}(\Delta t)$ is an overlap factor and $\rho_u$ is the instantaneous correlation of the two Brownian innovations in operational time. For independent Poisson refresh clocks, using the pooled rate in the continuous exponential-lag aggregation form gives the tractable approximation
\begin{equation}
\rho_{\Delta t}^{\rm sub} \simeq \rho_u f(\lambda_{12}^{\rm clk}\Delta t),
 \label{eq:pooled-clock-response}
\end{equation}
where the ordinary aggregation response is
\begin{equation}
f(x):=1-\frac{1-e^{-x}}{x}
 \label{eq:ordinary-aggregation-response}
\end{equation}for $x\geq0$, and $f(0):=0$. Since $\mathcal A_{12}(\Delta t)\to1$ as $\Delta t\to\infty$, the clock-only benchmark has $\rho_\infty=\rho_u$. This is analogous to the continuous exponential-lag aggregation envelope of \citet{toth2009}, but with $\lambda_{12}^{\rm clk}$ interpreted here as the effective clock-overlap rate.

See \ref{app:subordination-only} where, for independent Poisson refresh clocks with rates \(\lambda_1^{\rm clk}\) and \(\lambda_2^{\rm clk}\), the pooled refresh process is Poisson with rate $\lambda_{12}^{\rm clk}=\lambda_1^{\rm clk}+\lambda_2^{\rm clk}$.
Here the waiting time until either book refreshes is the minimum of two independent exponential waiting times and is therefore exponential with the pooled rate. The resulting overlap envelope remains an effective modelling approximation.

\subsection{Coupling-only mechanism}

Under the identity-clock benchmark and the normalized reaction-front projection with the local validity conditions in \ref{app:coupling-only}, the coupling gives the calendar-time log-price approximation
\begin{align}
 dP^{(1)}(t)&=\sigma_1 dB_1(t)-\kappa_{12}z_{12}(t)dt, \\
 dP^{(2)}(t)&=\sigma_2 dB_2(t)+\kappa_{21}z_{12}(t)dt,
 \label{eq:coupling-only}
\end{align}
where $z_{12}(t)=P^{(1)}(t)-P^{(2)}(t)$ and $\kappa_{12},\kappa_{21}>0$. The spread relaxes with rate $\kappa=\kappa_{12}+\kappa_{21}$.
Integrating the resulting cross-covariance kernel gives
\begin{equation}
\rho_{\Delta t}^{\rm coup} \simeq \rho_\infty f(\kappa\Delta t),
 \label{eq:coupling-response}
\end{equation}

This has the same aggregation-response form as \citet{toth2009}; see Equation~\eqref{eq:ordinary-aggregation-response}. In our derivation, its rate is now the reaction-front spread-relaxation rate $\kappa$.

For the closed two-log-price SDE model, the stationary spread implies unit long-aggregation correlation. In the normalised envelope above, \(\rho_\infty\) is retained as a phenomenological large-scale level for broader log-price dynamics outside that reduced closure. The ordered-pair coefficient \(\kappa_{jk}\) is the effective rate with which coupling-induced order flow from book \(k\) moves the reaction boundary of book \(j\). Since the two boundaries move in opposite directions when the spread is nonzero, both motions reduce the same spread.

\subsection{Combined mechanism}

When both effects are present, a leading-order separability approximation gives
\begin{equation}
\rho_{\Delta t}^{\rm comb} \simeq \rho_\infty\mathcal A_{12}(\Delta t)f(\kappa\Delta t).
 \label{eq:general-combined-correlation}
\end{equation}
For fractional clocks the exponential term is replaced by a Mittag--Leffler function \cite{Haubold2011MittagLeffler}. 
In the combined approximation, realised correlation emerges as the books sample overlapping operational time and the coupling-induced reaction-boundary response transmits the common movement. This product form depends on the separability assumption in \ref{app:combined-subordination-coupling}.

\section{Numerical implementation}
\label{sec:numerical-implementation}

The numerical implementation follows the same separation used in the analysis. Both order books are first evolved on a fixed operational-time and log-price lattice. The resulting reaction-boundary paths are then observed through separate calendar clocks. Calendar waiting times do not enter the density recurrence. 

\subsection{Operational-time evolution}

The operational grid is
\begin{equation}
    x_i=i\Delta x,
    \qquad
    u_n=n\Delta u,
\end{equation}
with fixed $\Delta x$ and $\Delta u$. At each step the two current density fields and their reaction boundaries are copied before the directed coupling terms are constructed. The additive sources and the two ordered coupling fields are assembled from this common pre-step state, and each book is advanced by the operational memory recurrence using those frozen inputs. The model assumes a unique simple reaction boundary; numerically, admissible exact zeros and strict sign crossings are recorded before the declared selection rule is applied.

This construction does not use event-dependent spatial displacements, off-grid neighbouring states or calendar-time interpolation. The numerical memory kernel remains part of the operational density evolution. The observation clocks enter only after the complete operational paths have been stored.

\subsection{Numerical reaction-front coupling}

For the coupling acting on book $j$ from book $k$, the current receiving-front translation mode is
\begin{equation}
    v_j^{\Delta x}(x_i,u_n)
    =-\partial_x^{\Delta x}\varphi^{(j)}(x_i,u_n),
\end{equation}
and the numerical coupling field is
\begin{equation}
    \ell_T^{(j,k)}(x_i,u_n)
    =-\kappa_{jk}z_{jk}(u_n)v_j^{\Delta x}(x_i,u_n).
    \label{eq:body-numerical-translation-source}
\end{equation}
The response rates $\kappa_{jk}$ are declared ordered boundary-response rates. They are not reconstructed during the simulation from the analytical regularisation parameters. This realises
\begin{equation}
    \dot p_j=-\kappa_{jk}(p_j-p_k),
    \qquad
    \dot z_{12}=-\bigl(\kappa_{12}+\kappa_{21}\bigr)z_{12},
\end{equation}
subject to the resolved-front and finite-step approximation. \ref{app:numerical-reaction-front-correspondence} gives the projection-level correspondence, while \ref{app:first-moment-correspondence} gives the analytical parameter-to-rate map. The analytical regularised source and Equation~\eqref{eq:body-numerical-translation-source} are not pointwise identical.

\subsection{Calendar-time observation}

Let $T_j(u_n)$ be the realised nondecreasing clock path for book $j$. Once the operational paths are complete, the calendar observation at time $t$ uses the previous completed operational state
\begin{equation}
    n_j(t)=\max\{n:T_j(u_n)\leq t\},
    \qquad
    P_\Delta^{(j)}(t)=p^{(j)}(u_{n_j(t)}).
    \label{eq:previous-refresh-map}
\end{equation}
The two books are evaluated on a common calendar grid within both realised clock horizons. Unsupported terminal observations are rejected. There is no extrapolation, interpolation or terminal-state clamping. Equation~\eqref{eq:previous-refresh-map} is the finite previous-refresh observation associated with the pathwise time-change algorithm in \ref{app:subordination-only}; it is not a non-uniform state update.

For two independent equal-rate Poisson clocks, the previous-refresh attenuation used for the numerical estimator has the individual book rate $\lambda^{\rm clk}$. Its clock-only reference is
\begin{equation}
 f(\lambda^{\rm clk}\Delta t).
 \label{eq:equal-rate-previous-refresh-reference}
\end{equation}
The sum $\lambda_1^{\rm clk}+\lambda_2^{\rm clk}$ is the rate of the minimum forward wait. It enters the effective pooled-rate approximation in Equation~\eqref{eq:pooled-clock-response}, not the equal-rate previous-refresh reference above.

\subsection{Clock, coupling and combined comparisons}

\begin{figure*}[htbp]
\centering
\includegraphics[width=\textwidth]{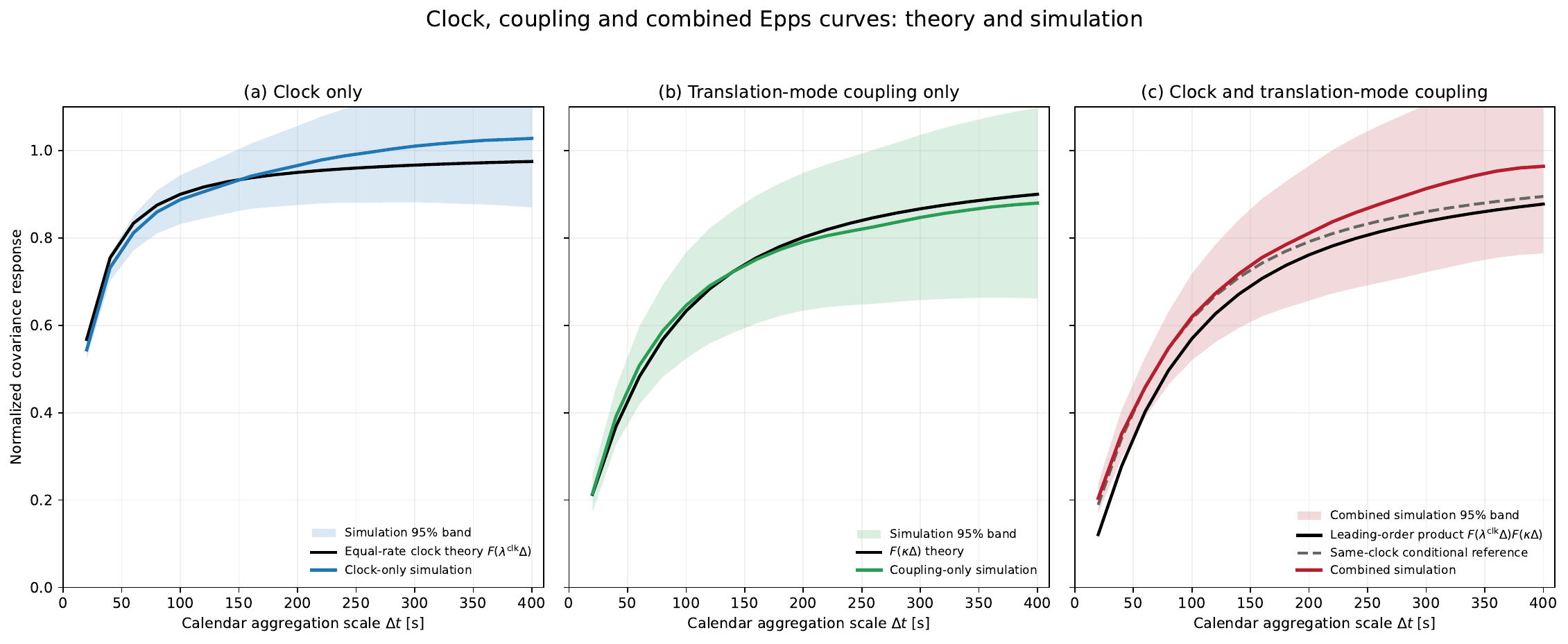}
\caption{\textbf{Clock, coupling and the emergence of correlation.} Here $f(x):=1-(1-e^{-x})/x$, Equation~\eqref{eq:ordinary-aggregation-response}. \textbf{Panel (a):} the black solid line is the clock-only curve for two independently observed books with the same refresh rate, $f(\lambda^{\rm clk}\Delta t)$, Equation~\eqref{eq:equal-rate-previous-refresh-reference}. Each calendar observation uses the last completed operational state, so short-scale correlation is reduced when the two books sample only partly overlapping operational-time intervals; see Equation~\eqref{eq:previous-refresh-map} and \ref{app:subordination-only}. \textbf{Panel (b):} the black solid line is the coupling-response factor $f(\kappa\Delta t)$, Equation~\eqref{eq:coupling-response}, with $\kappa=\kappa_{12}+\kappa_{21}$ as in Equation~\eqref{eq:numerical-spread-relaxation} and \ref{app:coupling-only}. \textbf{Panel (c):} the black solid line is the leading reference product $f(\lambda^{\rm clk}\Delta t)f(\kappa\Delta t)$, Equation~\eqref{eq:figure-leading-product}. The general combined approximation remains $\mathcal A_{12}(\Delta t)f(\kappa\Delta t)$ in Equation~\eqref{eq:general-combined-correlation}. The same-clock conditional curve retains the finite-step covariance on the operational intervals selected by the realised clocks; see \ref{app:combined-subordination-coupling}.}
\label{fig:final-estimator-aware-epps}
\end{figure*}

Figure~\ref{fig:final-estimator-aware-epps} separates the three numerical comparisons. Panel~(a) applies independent previous-refresh clocks to the completed thick reaction-boundary paths while excluding coupling. The comparison uses the equal per-book refresh rate $\lambda^{\rm clk}=0.1\,\mathrm{s}^{-1}$; the pooled minimum-forward-wait rate is a different object and is not substituted into the previous-refresh attenuation curve. The numerical correlation follows the clock-only reference closely, subject to the finite-front and previous-refresh construction in Equation~\eqref{eq:previous-refresh-map} and \ref{app:subordination-only}. 

Panel~(b) uses identity clocks and translation-mode coupling with $\kappa=0.025\,\mathrm{s}^{-1}$, and $\kappa_{12}=\kappa_{21}=0.0125\,\mathrm{s}^{-1}$.
The numerical covariance response is consistent with the prescribed reduced rate over the tested front and discretisation. The translation-mode implementation is given by Equation~\eqref{eq:body-numerical-translation-source}; its projection onto the reaction-boundary motion is described in \ref{app:numerical-reaction-front-correspondence}, with the analytical rate construction given in Appendices~\ref{app:coupling-only} and~\ref{app:first-moment-correspondence}. The comparison tests the reduced reaction-front response, not pointwise equality of the analytical and numerical source fields. 

Panel~(c) combines the clock and coupling components without refitting. The numerical comparison uses the leading reference product
\begin{equation}
 f(\lambda^{\rm clk}\Delta t)f(\kappa\Delta t),
 \label{eq:figure-leading-product}
\end{equation}
which specialises the clock factor to the equal-rate previous-refresh reference in Equation~\eqref{eq:equal-rate-previous-refresh-reference}. The general analytical approximation remains Equation~\eqref{eq:general-combined-correlation}, under the separability conditions in \ref{app:combined-subordination-coupling}.  The simulation is also compared with the finite-grid, finite-step conditional moment evaluated on the operational intervals selected by the same realised clocks. This same-clock reference gives the closer finite-grid comparison, without refitting any clock rate, coupling rate, boundary normalisation, time map or simulation parameter. 

The operational and observation algorithms, numerical diagnostics and machine-readable outputs used for these comparisons are supplied with the reproducibility archive \citep{AngstmannGebbie2026ReproducibilityCode}.

\section{Discussion}
\label{sec:discussion}

We connect three mature, but largely disjoint, parts of the market-microstructure literature: reaction--diffusion order books, reduced-form accounts of the Epps effect, and discrete-event models of trading. Each is well developed in its own right.

Reaction--diffusion models successfully capture liquidity replenishment and price impact in single assets, but typically remain uncoupled across assets and make no claims about cross--correlations \citep{donier2015,toth2011,benzaquen2018}. Conversely, Epps--effect models explain scale--dependent correlation using asynchronous sampling, lead--lag effects, or estimator bias, but do so at the level of observed prices rather than order--flow dynamics \citep{epps1979,lo1990,hayashi2005,mastromatteo2011}.
We connect these strands by embedding correlation emergence directly into a coupled reaction--diffusion representation of multiple order books.

Correlation emerges from explicit order-flow coupling between the books rather than from correlated price noise or a latent multivariate diffusion. A limit-order-book Hawkes model of trade event types does not directly specify a scalar log-return correlation; events must first be mapped into price changes. The cross-dependence is nevertheless structural in the latent event model~\footnote{For example, in the four-dimensional Hawkes price model of \citet{chang2025discrete}, the cross-kernel
$\phi^{(c)}(t) =\alpha^{(c)}e^{-\beta t}\mathbf{1}_{\{t>0\}}$
connects price-move events across the two assets. The co-dependence is not strictly emergent, nor phenomenological.}. Our model places the cross-book mechanism in $\gamma_{jk}$, with its price effect transmitted through the coupled order-density fields and moving reaction fronts and as such is sourced from the phenomenology of trading itself rather than being latent pre-specified relationships between trade or event types.

This is also why Hawkes estimation is conditional on the market implementation: matching-engine and order-handling rules can distort parameter estimates and model identification \citep{JericevichChangGebbie2021}. In addition, structural inference in an open and reflexive financial market remains conditional on any latent model imposed on the data \citep{PolakowGebbieFlint2026} irrespective of the calibration approach. Hawkes estimation remains valid, but an estimated excitation network is not automatically the recovered mechanics of the market. This qualification matters when structural and phenomenological parameters are compared.

The reaction boundary carries the correlation between the books. In the reaction--diffusion model, the log-mid-price is defined implicitly as a moving annihilation boundary between buy and sell densities. By analysing how the coupling perturbs the order density near this boundary, we can show that correlation transmission is filtered by the finite relaxation time of the reaction front. This provides a microstructural explanation for the Epps effect that we think is absent from earlier reduced--form derivations, where correlation attenuation is attributed solely to sampling effects or lead--lag structure \citep{toth2009,chang2021}.
Even perfectly synchronous observation does not eliminate short--horizon correlation decay when the coupling response is slow. 

\citet{toth2009} obtain the same elementary exponential aggregation envelope by aggregating exponentially decaying lagged correlations and relate its characteristic scale to the reaction time of market participants. Here $\lambda_{12}^{\rm clk}$ is an event-clock rate, whereas $\kappa$ is a reaction-front spread-relaxation rate derived from the coupled order-book dynamics. This separates the clock and coupling mechanisms and extends the treatment to the fractional and qualified combined cases.

One caution is that the one-front construction treats the zero crossing as the model log-mid-price. Exponentiating it gives the corresponding model mid-price level and, at this reduced level, an approximation to the transaction price. Its finite width coarse-grains discrete log-price levels and part of the unresolved within-book spread structure, but it does not explicitly resolve best bid and best ask. That extension would require two moving reaction fronts together with nontrivial buy- and sell-side currents. 

Nevertheless, this construction allows us to clarify the relation between discrete event dynamics and their continuum limits. Concretely, the exponent $\alpha_u$ belongs to the operational-time density law, while $\alpha_t$ belongs to the calendar-time observation clock. The restriction $\alpha_u=\alpha_t$ is imposed only when the two waiting-time descriptions are deliberately identified \citep{montroll1965,scalas2006,angstmann2016}.

Separating operational time from the observation clocks distinguishes clock overlap from coupling response and their combined effect. Similar Epps envelopes can therefore have different sources. Earlier work separated asynchrony from lead--lag effects \citep{mastromatteo2011}; our treatment starts from interacting order books and keeps the clock and coupling parameters distinct.

The numerical comparisons preserve the same mechanism separation. The clock-only experiment isolates previous-refresh attenuation, while the identity-clock experiment isolates reaction-front coupling. Their combination is then a holdout calculation with the component rates and normalisation left unchanged. The leading product remains the analytical mechanism decomposition. The same-clock conditional reference instead resolves the finite estimator on the realised clock paths, which is why it can improve the finite-sample comparison without replacing the leading-order theory.

The analytical and numerical coupling fields have different roles. The regularised source is a local closure used to derive the ordered reaction-front response while the numerical field applies that response rate directly to the current receiving front. Their correspondence is therefore projection-based rather than pointwise. The two source fields differ, but they generate the same directed mean-reverting boundary response after being projected. The key point made here is that both the analytic treatment and the numerical treatement separate the finite-response from the clock subordinations. 

The numerical comparison can be considered limited in the sense that it tests the rate recovery for the front and its discretisation but it does not establish analytic correspondence, or pointwise closure. This analysis complements arguments that the Epps effect reflects the discrete-event nature of markets \citep{chang2025discrete} by showing how that discreteness appears after the diffusion limit and which additional assumptions are required.

\section{Conclusion}

We obtain an analytical explanation of correlation emergence and the Epps effect from the discrete-to-continuum construction of coupled order books and a regularised representation of the pair-trader coupling. Separating reaction-boundary dynamics in operational time from the asynchronous observation clocks distinguishes attenuation due to incomplete clock overlap from attenuation due to the finite coupling response. This provides an analytical counterpart and explanation to the prior numerical simulation studies \citep{BauerDianaGebbie2024,DianaGebbie2025} which used a non-uniformly sampled grid implementation were the time mixing with the finite coupling response could not easily be separated. 

The numerical implementation here now directly follows the current separation. The two books evolve on a common operational lattice, the ordered coupling acts through the receiving reaction front, and the completed boundary paths are observed through separate previous-refresh clocks. Figure \ref{fig:final-estimator-aware-epps} shows that the component calculations recover the clock and coupling responses over the tested numerical regime. In the combined experiment, the finite-step conditional reference improves on the leading product while leaving the mechanism decomposition intact. The remaining difference are thus finite-front, finite-step, and estimator-level effects.

\section*{Code and data availability}

The code and numerical outputs supporting this paper are available from the University of Cape Town reproducibility archive \citep{AngstmannGebbie2026ReproducibilityCode}, \href{https://doi.org/10.25375/uct.33368986.v1}{\texttt{doi:10.25375/uct.33368986.v1}}. The corresponding public GitHub repository is available at \url{https://github.com/timgebbie/correlation-emergence-reproducibility/releases/tag/v2.0.0}. No empirical market data are used.

\section*{Acknowledgements}

We acknowledge conversations with Derick Diana, Byron Jacobs and Dominic Bauer.

\section*{Declaration of generative AI usage} 
During manuscript revision, the authors used ChatGPT and Microsoft Copilot for proofreading, consistency checking and notation review. All mathematical arguments, scientific claims and final wording were checked and approved by the authors, who take full responsibility for the content of the article.

\bibliography{CoupledOB-v4.0.0}

\appendix

\section{Clock subordination}
\label{app:subordination-only}

This appendix derives an Epps-effect curve generated only by non-uniform observation clocks. The operational-time order-density dynamics is defined before the clock is applied. The pair-trader coupling is excluded. The observation clocks introduced here are analytical objects used to isolate calendar-time overlap attenuation. 

\subsection{Time change}

Let $\varphi^{(j)}(x,u)$ solve the operational-time order-density equation and let
\begin{equation}
    p^{(j)}(u)=\{x:\varphi^{(j)}(x,u)=0\}
\end{equation}
be its unique reaction boundary. For an increasing process $T_j(u)$, define the inverse clock
\begin{equation}
    E_j(t)=\inf\{u\geq0:T_j(u)>t\}.
\end{equation}
The calendar-time field is
\begin{equation}
    \Phi^{(j)}(x,t)=\varphi^{(j)}\!\left(x,E_j(t)\right).
\end{equation}
Uniqueness of the zero crossing gives
\begin{equation}
    P^{(j)}(t)=\{x:\Phi^{(j)}(x,t)=0\}=p^{(j)}\!\left(E_j(t)\right).
\end{equation}
This is the subordination used in the body. A closed calendar-time equation for $\Phi^{(j)}$ follows without further closure for a linear operational-time generator. The nonlinear coupled field is defined pathwise by the time change; a closed fractional PDE requires an additional approximation.

\subsection{Log-price assumptions}

We use the following assumptions.
\begin{enumerate}
\item[(A1)] A filtered probability space supports the two operational-time Brownian innovations, with instantaneous correlation $\rho_u$.
\item[(A2)] The reaction-boundary log-prices are approximated locally by continuous operational-time semimartingales. Drifts are omitted because their short-window covariance contribution is lower order than the martingale covariance.
\item[(A3)] The inverse observation clocks $E_j$ are nondecreasing, right-continuous and exogenous to the Brownian innovations. Their increments have finite first moments. Stationary increments are assumed only when a stationary expression is used.
\item[(A4)] Conditional on the clocks, the covariance of the two observed increments is determined by the overlap of the sampled operational-time intervals.
\end{enumerate}

\subsection{Operational-time reduction}

For the subordination-only calculation, approximate the two reaction-boundary log-prices in a common operational time by
\begin{align}
    dp^{(1)}(u) &= \sigma_1dB^{(1)}(u),\\
    dp^{(2)}(u) &= \sigma_2dB^{(2)}(u),
\end{align}
with
\begin{equation}
    d\langle B^{(1)},B^{(2)}\rangle_u=\rho_u\,du,
\end{equation}
where $\rho_u$ is the operational-time innovation correlation. For the conditional variance and covariance calculations below, the observation clocks are assumed exogenous to the operational-time Brownian innovations. Calendar-time log-prices are $P^{(j)}(t)=p^{(j)}(E_j(t))$. For $I=(t,t+\Delta t]$, define the sampled operational-time intervals
\begin{equation}
    J_j(I)=\big(E_j(t),E_j(t+\Delta t)\big]
\end{equation}
and their overlap
\begin{equation}
    \Theta_{12}(I)=\left|J_1(I)\cap J_2(I)\right|.
\end{equation}
Also write $\Delta E_j(I)=E_j(t+\Delta t)-E_j(t)$ and $R_{\Delta t}^{(j)}=P^{(j)}(t+\Delta t)-P^{(j)}(t)$. Conditional on the clocks,
\begin{equation}
    R_{\Delta t}^{(j)}
    =\sigma_j\int_{E_j(t)}^{E_j(t+\Delta t)}dB^{(j)}(u).
\end{equation}
Writing these stochastic integrals on the common operational-time axis gives
\begin{align}
    R_{\Delta t}^{(1)}
    &=\sigma_1\int_0^\infty
    \mathbf 1_{J_1(I)}(u)dB^{(1)}(u),\\
    R_{\Delta t}^{(2)}
    &=\sigma_2\int_0^\infty
    \mathbf 1_{J_2(I)}(u)dB^{(2)}(u).
\end{align}
The It\^{o} isometry then gives
\begin{equation}
    \operatorname{Var}(R_{\Delta t}^{(j)}\mid E_j)=\sigma_j^2\Delta E_j(I),
\end{equation}
while
\begin{equation}
    \operatorname{Cov}(R_{\Delta t}^{(1)},R_{\Delta t}^{(2)}\mid E_1,E_2)
    =\rho_u\sigma_1\sigma_2\Theta_{12}(I).
\end{equation}
Taking expectations gives
\begin{equation}
    \rho_{\Delta t}^{\rm sub}=\rho_u\mathcal A_{12}(\Delta t),
\end{equation}
where
\begin{equation}
    \mathcal A_{12}(\Delta t)
    =\frac{\mathbb E[\Theta_{12}(I)]}
    {\sqrt{\mathbb E[\Delta E_1(I)]\mathbb E[\Delta E_2(I)]}}.
\end{equation}
Here $t$ is fixed and suppressed in the notation. Since inverse-clock increments need not be stationary, the overlap factor may also depend on $t$.
If $E_1=E_2$, then $\Theta_{12}(I)=\Delta E_1(I)$ and there is no separate attenuation from incomplete clock overlap.

\subsection{Poisson clocks}

If the books refresh independently with effective rates $\lambda_1^{\rm clk}$ and $\lambda_2^{\rm clk}$, the waiting time until either book refreshes is exponential with the pooled rate
\begin{equation}
    \lambda_{12}^{\rm clk}=\lambda_1^{\rm clk}+\lambda_2^{\rm clk}.
\end{equation}
The probability that neither book refreshes over a lag $s$ is
\begin{equation}
    \mathbb P(\text{no refresh of either book over lag }s)
    =e^{-\lambda_{12}^{\rm clk}s}.
\end{equation}
Averaging the lag-dependent overlap over the aggregation window gives
\begin{align}
    \mathcal A_{12}^{\rm Pois}(\Delta t)
    &\simeq
    \frac{1}{\Delta t}\int_0^{\Delta t}
    \left(1-e^{-\lambda_{12}^{\rm clk}s}\right)ds\\
    &=1-\frac{1}{\Delta t}\int_0^{\Delta t}
    e^{-\lambda_{12}^{\rm clk}s}ds\\
    &=1-\frac{1-e^{-\lambda_{12}^{\rm clk}\Delta t}}
    {\lambda_{12}^{\rm clk}\Delta t}
    =f(\lambda_{12}^{\rm clk}\Delta t).
\end{align}
Therefore
\begin{equation}
 \rho_{\Delta t}^{\rm sub}\simeq\rho_u f(\lambda_{12}^{\rm clk}\Delta t),
\end{equation}
where $f$ is defined in Equation~\eqref{eq:ordinary-aggregation-response}.
The factor is asymptotically $\lambda_{12}^{\rm clk}\Delta t/2$ as $\Delta t\downarrow0$ and tends to one as $\Delta t\to\infty$. Hence $\rho_\infty=\rho_u$ in this benchmark. The pooled-refresh rate is exact under independent Poisson waiting times; the overlap envelope remains an effective modelling approximation.

\subsection{Fractional clocks}

For an inverse-stable clock of order $0<\alpha_t\leq1$, the exponential survival term is replaced by a Mittag--Leffler survival term. If the operational-time log-price is Brownian, then
\begin{equation}
    \operatorname{Var}[P^{(j)}(t)-P^{(j)}(0)]
    =\sigma_j^2\mathbb E[E_j(t)]
    =\frac{\sigma_j^2t^{\alpha_t}}{\Gamma(1+\alpha_t)},
    \label{eq:fractional-price-variance}
\end{equation}
for the standard inverse-stable normalization. Ordinary calendar-time log-price diffusion is recovered at $\alpha_t=1$. Under this log-price reduction, $0<\alpha_t<1$ gives sublinear variance growth. This result follows from the log-price clock, not from the operational-time density equation alone.

Introduce a fractional clock scale $\lambda_{12,\alpha_t}^{\rm clk}$ with units of inverse time to the power $\alpha_t$ and $\left.\lambda_{12,\alpha_t}^{\rm clk}\right|_{\alpha_t=1}=\lambda_{12}^{\rm clk}$. Then
\begin{equation}
    E_{\alpha_t}(-\lambda s^{\alpha_t})
    =\sum_{n=0}^{\infty}
    \frac{(-\lambda)^ns^{\alpha_t n}}
    {\Gamma(\alpha_t n+1)}.
\end{equation}
Termwise integration gives
\begin{align}
    \frac{1}{\Delta t}\int_0^{\Delta t}
    E_{\alpha_t}(-\lambda s^{\alpha_t})ds
    &=\sum_{n=0}^{\infty}
    \frac{(-\lambda)^n(\Delta t)^{\alpha_t n}}
    {(\alpha_t n+1)\Gamma(\alpha_t n+1)}\\
    &=E_{\alpha_t,2}(-\lambda(\Delta t)^{\alpha_t}).
\end{align}
Hence
\begin{equation}
    \mathcal A_{12}^{(\alpha_t)}(\Delta t)
    \simeq
    1-E_{\alpha_t,2}(-\lambda_{12,\alpha_t}^{\rm clk}(\Delta t)^{\alpha_t}).
\end{equation}
Since $E_{1,2}(-x)=(1-e^{-x})/x$, this reduces to the Poisson-refresh expression at $\alpha_t=1$. The Mittag--Leffler expression is an effective fractional extension of the exponential overlap envelope, not an exact joint-overlap law for arbitrary inverse-stable clocks. The exponent $\alpha_t$ is calibrated from the observation-clock waiting times or calendar-time anomalous scaling. It is not fixed by $\alpha_u$.

\section{Coupling response}
\label{app:coupling-only}

This appendix derives an approximate Epps curve generated by the finite response time of the regularised pair-trader coupling and states the validity conditions for the reaction-front reduction. The front calculation is carried out in operational time. The coupling-only Epps curve then uses the identity clock $E_j(t)=t$.

\subsection{Assumptions}

We use the following assumptions.
\begin{enumerate}
\item[(B1)] Each signed order-density field is differentiable in operational time and twice differentiable in log-price near its reaction boundary for the deterministic local calculation.
\item[(B2)] Each model log-mid-price is a unique simple zero of its signed order-density field.
\item[(B3)] The pair-trader source is local near the reaction boundary and has a finite first spatial moment.
\item[(B4)] The coupling is weak enough that the front profile and its response mode may be frozen over the short coupling-response calculation.
\item[(B5)] The Brownian terms introduced after the deterministic projection are effective log-price innovations. They are not obtained by applying the ordinary chain rule to the complete stochastic density field.
\end{enumerate}

\subsection{Reaction-front reduction}

The model log-mid-price of book $j$ is the unique zero crossing
\begin{equation}
    \varphi^{(j)}(p^{(j)}(u),u)=0.
\end{equation}
Assume that the locally frozen density field is differentiable near a simple zero and consider the deterministic finite-variation displacement induced by the coupling. Writing
\begin{equation}
    \mathcal L_j(u)=\partial_x\varphi^{(j)}(p^{(j)}(u),u)\neq0,
\end{equation}
differentiation by the ordinary chain rule gives
\begin{equation}
    \frac{d}{du}p_{\rm coup}^{(j)}(u)
    =-\frac{\left.\partial_u\varphi^{(j)}(p^{(j)}(u),u)\right|_{\rm coup}}
    {\mathcal L_j(u)}.
\end{equation}
This identity is used only for the deterministic coupling-induced displacement of the locally frozen front. It is not asserted as a chain rule for the complete stochastic density field. A stochastic moving boundary would generally require an It\^{o}--Wentzell treatment. The Brownian innovations introduced below are therefore effective log-price terms and are not derived from this identity. The identity depends on the point value of the source. The regularised source is odd locally and satisfies
\begin{equation}
    \ell_\varepsilon^{(j,k)}(p^{(j)}(u),u)=0.
\end{equation}
Its distributed effect on a finite-width reaction front therefore requires a separate collective-coordinate closure; it does not follow from point evaluation alone.

Let $\varphi_j^\star(y)$ be the locally frozen front profile, where $y=x-p^{(j)}(u)$, and define its translational mode by
\begin{equation}
    v_j(y)=-\partial_y\varphi_j^\star(y).
\end{equation}
If $\psi_j$ is the corresponding adjoint response mode, with inner product $\langle\cdot,\cdot\rangle$, set
\begin{equation}
    N_j=\langle\psi_j,v_j\rangle\neq0.
\end{equation}
Projection of the locally linearised density equation onto $\psi_j$ gives the normalized coupling-induced velocity
\begin{equation}
    \frac{d}{du}p_{\rm coup}^{(j)}(u)
    =\frac{\langle\psi_j,\ell_\varepsilon^{(j,k)}\rangle}{N_j}.
    \label{eq:normalised-front-projection}
\end{equation}
The sign is fixed by the definition of $v_j$. If, over the active support of the coupling source,
\begin{equation}
    \frac{\psi_j(y)}{N_j}\simeq A_jy,
    \label{eq:first-moment-mode}
\end{equation}
then Equation~\eqref{eq:normalised-front-projection} reduces to the first-moment closure
\begin{equation}
    \frac{d}{du}p_{\rm coup}^{(j)}(u)
    \simeq A_j\int_{\mathbb R}y\ell_\varepsilon^{(j,k)}(y,z_{jk})\,dy.
    \label{eq:first-moment-front-closure}
\end{equation}
The coefficient $A_j$ contains the normalization, orientation and dimensions required by the projection. If $\varphi^{(j)}$ is measured as signed order volume per unit log-price, then $A_j$ has units of inverse order volume, so Equation~\eqref{eq:first-moment-front-closure} has units of log-price per unit operational time. \ref{app:first-moment-correspondence} proves that the raw moment is $\gamma_{jk}M_j^+z_{jk}$, where
\begin{align}
    M_j^+&=\int_0^\infty yq_j(y)\,dy
    &=-\lambda_j\mu_j\int_0^\infty
    y^2e^{-\mu_jy^2}dy\\
    &=-\frac{\lambda_j\sqrt{\pi}}{4\sqrt{\mu_j}},
\end{align}
using
\begin{equation}
    \int_0^\infty y^2e^{-\mu y^2}dy
    =\frac{\sqrt{\pi}}{4\mu^{3/2}}.
\end{equation}
Thus
\begin{align}
    \frac{d}{du}p_{\rm coup}^{(j)}
    &\simeq A_j\gamma_{jk}M_j^+z_{jk}
    =-\kappa_{jk}z_{jk},\\
    \kappa_{jk}&:=-A_j\gamma_{jk}M_j^+>0.
    \label{eq:ordered-pair-rate}
\end{align}
For the signed-density convention used here we assume $A_j>0$. Since $M_j^+<0$ and $\gamma_{jk}>0$, this gives the positive mean-reverting rate $\kappa_{jk}>0$.

\subsection{Linear response}

Let $z(u)=z_{12}(u)=p^{(1)}(u)-p^{(2)}(u)$. Around $z=0$, the side-selected coupling induces a stabilising linear spread response. The operational-time log-price approximation is
\begin{align}
    dp^{(1)}(u) &= \sigma_1dB_1(u)-\kappa_{12}z(u)du,\\
    dp^{(2)}(u) &= \sigma_2dB_2(u)+\kappa_{21}z(u)du,
\end{align}
where $B_1$ and $B_2$ are independent Brownian motions. Hence
\begin{equation}
    dz(u)=-(\kappa_{12}+\kappa_{21})z(u)du+\sigma_zdB_z(u),
\end{equation}
where
\begin{equation}
    \sigma_z^2=\sigma_1^2+\sigma_2^2,
    \qquad
    \sigma_zdB_z=\sigma_1dB_1-\sigma_2dB_2,
\end{equation}
and $\kappa=\kappa_{12}+\kappa_{21}$ is the operational-time coupling response rate. Thus $B_z$ is constructed from the two log-price innovations rather than introduced as a third independent source. For the coupling-only calculation below, the identity clock identifies $u$ with calendar time $t$.

\subsection{Validity}

For a coupling-induced front displacement $\xi_j$, write
\begin{equation}
    \varphi_j^\star(\xi_j)
    =\mathcal L_j\xi_j
    +\frac12\mathcal C_j\xi_j^2
    +O(\xi_j^3),
    \qquad
    \mathcal C_j=\partial_y^2\varphi_j^\star(0).
\end{equation}
The Taylor remainder is small relative to the retained term when
\begin{equation}
    \eta_j^{\rm disp}
    :=\frac{|\mathcal C_j|\,|\xi_j|}{2|\mathcal L_j|}
    \ll1.
\end{equation}
The simple-root condition $\mathcal L_j\neq0$ must hold, and the coupling-source width $w_{q,j}\sim\mu_j^{-1/2}$ must remain within the locally linear part of the front:
\begin{equation}
    w_{q,j}\ll L_{{\rm curv},j}.
\end{equation}
The frozen-shape approximation also requires the front-shape evolution time to exceed the response time,
\begin{equation}
    \tau_{{\rm shape},j}\gg\kappa^{-1}.
\end{equation}
The reduction may fail for a flat or non-simple zero, multiple zero crossings, large shocks or displacements, strong coupling that deforms rather than translates the front, a source broad relative to the local curvature scale, front evolution on the response timescale, or asymmetric finite-domain and grid effects. Outside this regime the response is more generally
\begin{equation}
    dp_{\rm coup}^{(j)}=b_j(z,u)du,
\end{equation}   
where $b_j(z,u)=-\kappa_{jk}z$ + higher-order + state-dependent terms. The spread is then not generally Ornstein--Uhlenbeck and the exponential response kernel below does not follow directly.

The finite one-front width coarse-grains discrete log-prices and part of the unresolved within-book spread structure. It does not guarantee local linearity or constitute an explicit bid--ask spread. Resolving that spread requires distinct bid and ask fronts with nontrivial buy- and sell-side currents.

\subsection{Covariance response}

For the closed two-log-price SDE, define
\begin{equation}
    C
    =\frac{\kappa_{12}^2\sigma_2^2+\kappa_{21}^2\sigma_1^2}
    {(\kappa_{12}+\kappa_{21})^2},
    \qquad
    A_{\Delta t}
    =\Delta t-\frac{1-e^{-\kappa\Delta t}}{\kappa}.
\end{equation}
The finite response time $\kappa^{-1}$ gives the even component of the lagged cross-covariance kernel
\begin{equation}
    c_{12}^{\rm even}(u)=C\frac{\kappa}{2}e^{-\kappa|u|}.
\end{equation}
A parameter-asymmetric model may also contain an odd lead--lag component. It vanishes under the balance condition $\kappa_{21}\sigma_1^2=\kappa_{12}\sigma_2^2$. Because the equal-window aggregation weight is even, the odd component does not contribute to the covariance calculated below.
The covariance of log returns over a window of length $\Delta t$ is
\begin{align}
    \operatorname{Cov}_{\Delta t}
    &=\int_{-\Delta t}^{\Delta t}
    (\Delta t-|u|)c_{12}^{\rm even}(u)du\\
    &=C\kappa\int_0^{\Delta t}
    (\Delta t-u)e^{-\kappa u}du.
\end{align}
The remaining integral is
\begin{align}
    \int_0^{\Delta t}(\Delta t-u)e^{-\kappa u}du
    &=\frac{\Delta t}{\kappa}
    -\frac{1-e^{-\kappa\Delta t}}{\kappa^2},
\end{align}
and therefore
\begin{equation}
    \operatorname{Cov}_{\Delta t}
    =C\left[\Delta t-
    \frac{1-e^{-\kappa\Delta t}}{\kappa}\right]
    =CA_{\Delta t}.
\end{equation}
The exact equal-window log-return variances of the closed SDE are
\begin{equation}
    \operatorname{Var}(R_{\Delta t}^{(j)})
    =\sigma_j^2\Delta t+(C-\sigma_j^2)A_{\Delta t}.
\end{equation}
Its exact correlation is therefore
\begin{equation*}
    \rho_{\Delta t}^{\rm SDE}
    =\frac{CA_{\Delta t}}
    {\sqrt{[\sigma_1^2\Delta t+(C-\sigma_1^2)A_{\Delta t}]
    [\sigma_2^2\Delta t+(C-\sigma_2^2)A_{\Delta t}]}}.
\end{equation*}
At short scales,
\begin{equation}
    \rho_{\Delta t}^{\rm SDE}
    \sim\frac{C\kappa}{2\sigma_1\sigma_2}\Delta t,
    \qquad \Delta t\downarrow0,
\end{equation}
while the stationary spread implies $\rho_{\Delta t}^{\rm SDE}\to1$ as $\Delta t\to\infty$.

For comparison with a broader log-price process containing variance components outside this reduced coupling closure, use the phenomenological approximation $\operatorname{Var}(R_{\Delta t}^{(j)})\simeq v_j\Delta t$. This gives
\begin{equation}
    \rho_{\Delta t}^{\rm coup}
    \simeq
    \rho_\infty f(\kappa\Delta t),
    \qquad
    \rho_\infty=\frac{C}{\sqrt{v_1v_2}}.
\end{equation}
Here $\rho_\infty$ is an effective large-scale normalisation rather than a free parameter of the closed two-log-price SDE. Setting $\rho_\infty=1$ recovers the long-scale implication of that closed model. The phenomenological expression is linear at the origin and converges to $\rho_\infty$ at large aggregation scales.

\section{Combined mechanism}
\label{app:combined-subordination-coupling}

This appendix combines the observation-clock and finite-coupling mechanisms. The analytical product separates calendar-time clock overlap from operational-time coupling response. It is a leading-order decomposition, not a pathwise identity: the factorisation requires the clock-overlap and local coupling-response terms to separate in expectation. In operational time the coupling produces an exponential response, while the inverse clock maps that response into calendar time. For inverse-stable subordination of order $\alpha_t$, the response becomes Mittag--Leffler, as in the theory of inverse stable subordinators \citep{MeerschaertStraka2013}.

\subsection{Assumptions}

We use the following assumptions.
\begin{enumerate}
\item[(C1)] In operational time, the coupling-induced covariance response has a completely monotone relaxation survival function $h(u)$ with response density $K(u)=-h'(u)$.
\item[(C2)] The observation clocks are exogenous to the effective log-price innovations and independent of the local coupling response to first order.
\item[(C3)] Variance growth is approximately linear over the aggregation scales used in the reduced expression. Otherwise the corresponding variance normalization must be retained explicitly.
\item[(C4)] The product formula is a leading-order separable approximation. It need not hold when the clock changes the local coupling intensity or pair-trader activity is driven by the same events that generate the observations.
\end{enumerate}

\subsection{Clock and coupling}

Let $H_\kappa(I)$ denote the normalized coupling response over the aggregation window. The product form requires the leading-order separability condition
\begin{equation}
    \mathbb{E}[\Theta_{12}(I)H_\kappa(I)]
    \simeq
    \mathbb{E}[\Theta_{12}(I)]\,\mathbb{E}[H_\kappa(I)].
    \label{eq:clock-coupling-separability}
\end{equation}
If the clocks mainly attenuate the operational-time overlap while the coupling response remains approximately exponential in calendar time, this gives
\begin{equation}
    \rho_{\Delta t}^{\rm comb}
    \simeq
    \rho_\infty\mathcal{A}_{12}(\Delta t)f(\kappa\Delta t).
\end{equation}
This form is useful for light-tailed non-uniform clocks because the two attenuation mechanisms remain separately identifiable. If the observation clocks are common, $E_1=E_2$, then $\mathcal A_{12}(\Delta t)=1$ and the expression reduces to the coupling-response curve. For a common observation clock, the asynchronous attenuation associated with distinct book-specific clocks disappears. Any temporal non-uniformity then belongs to the common observation map rather than to the operational state update, which remains defined on the fixed uniform operational lattice. With distinct observation clocks, $E_1\ne E_2$, the additional overlap factor remains nontrivial. If clock overlap and coupling transmission are materially dependent, the expectation does not factorise and the product form need not hold.

\subsection{Fractional response}

Let $h(u)$ denote the operational-time relaxation survival function. Ordinary relaxation solves
\begin{equation}
    \frac{d}{du}h(u)=-\kappa h(u),
    \qquad h(0)=1,
\end{equation}
with $h(u)=e^{-\kappa u}$. Under inverse-stable subordination of order $0<\alpha_t\leq1$, introduce a calendar-time response scale $\kappa_{\alpha_t}$ with units of inverse time to the power $\alpha_t$ and $\left.\kappa_{\alpha_t}\right|_{\alpha_t=1}=\kappa$. The subordinated scalar relaxation uses the Caputo derivative:
\begin{equation}
    {}^{\mathrm C}D_t^{\alpha_t}h_{\alpha_t}(t)
    =-\kappa_{\alpha_t}h_{\alpha_t}(t),
    \qquad h_{\alpha_t}(0)=1.
\end{equation}
For $0<\alpha_t<1$,
\begin{equation}
    {}^{\mathrm C}D_t^{\alpha_t}h(t)
    =\frac{1}{\Gamma(1-\alpha_t)}
    \int_0^t\frac{h'(s)}{(t-s)^{\alpha_t}}\,ds,
\end{equation}
Taking Laplace transforms of the fractional relaxation equation gives
\begin{equation}
    s^{\alpha_t}\widehat h_{\alpha_t}(s)-s^{\alpha_t-1}
    =-\kappa_{\alpha_t}\widehat h_{\alpha_t}(s),
\end{equation}
and hence
\begin{equation}
    \widehat h_{\alpha_t}(s)
    =\frac{s^{\alpha_t-1}}
    {s^{\alpha_t}+\kappa_{\alpha_t}}.
\end{equation}
Inverting gives
\begin{equation}
    h_{\alpha_t}(t)=E_{\alpha_t}(-\kappa_{\alpha_t}t^{\alpha_t}).
\end{equation}
The corresponding response density is
\begin{align}
    K_{\alpha_t,\kappa_{\alpha_t}}(t)
    &=-\frac{d}{dt}E_{\alpha_t}
    (-\kappa_{\alpha_t}t^{\alpha_t})\\
    &=\kappa_{\alpha_t}t^{\alpha_t-1}
    E_{\alpha_t,\alpha_t}
    (-\kappa_{\alpha_t}t^{\alpha_t}).
\end{align}
Since $K=-h'$, integration by parts gives
\begin{align}
    \int_0^{\Delta t}(\Delta t-s)K(s)ds
    &=\Delta t-\int_0^{\Delta t}h(s)ds.
\end{align}
The normalised covariance response is
\begin{align}
    F_{\alpha_t}(\Delta t)
    &=\frac{1}{\Delta t}\int_0^{\Delta t}(\Delta t-s)
    K_{\alpha_t,\kappa_{\alpha_t}}(s)ds\\
    &=1-\frac{1}{\Delta t}\int_0^{\Delta t}
    E_{\alpha_t}(-\kappa_{\alpha_t}s^{\alpha_t})ds\\
    &=1-E_{\alpha_t,2}(-\kappa_{\alpha_t}(\Delta t)^{\alpha_t}).
\end{align}
Therefore
\begin{equation}
    \rho_{\Delta t}^{(\alpha_t)}
    \simeq
    \rho_\infty\left[1-E_{\alpha_t,2}
    (-\kappa_{\alpha_t}(\Delta t)^{\alpha_t})\right].
\end{equation}
For $\alpha_t=1$, $E_{1,2}(-x)=(1-e^{-x})/x$, recovering the coupling-only curve. For small $\Delta t$,
\begin{equation}
    E_{\alpha_t,2}(-\kappa_{\alpha_t}(\Delta t)^{\alpha_t})
    =1-\frac{\kappa_{\alpha_t}(\Delta t)^{\alpha_t}}
    {\Gamma(\alpha_t+2)}
    +O((\Delta t)^{2\alpha_t}),
\end{equation}
so
\begin{equation}
    \rho_{\Delta t}^{(\alpha_t)}
    \sim
    \rho_\infty\frac{\kappa_{\alpha_t}(\Delta t)^{\alpha_t}}
    {\Gamma(\alpha_t+2)}.
\end{equation}
The inverse clock changes the short-scale emergence of correlation from linear order in $\Delta t$ to order $(\Delta t)^{\alpha_t}$. Identifying $\alpha_t$ with the operational-time density exponent $\alpha_u$ is an additional modelling restriction.

\section{Numerical correspondence}\label{app:numerical-reaction-front-correspondence}

The analytical regularisation and the numerical coupling have different roles. The analytical construction introduces a bounded local source and projects it onto the displacement mode of the receiving reaction front. The numerical implementation starts from the reduced ordered response rate and acts directly through that receiving-front translation mode. The correspondence is therefore at the level of the reaction-boundary motion, not pointwise equality of the two source fields.

\subsection{Analytical projection}

Let
\begin{equation}
    y_j=x-p^{(j)}(u),
    \qquad
    z_{jk}(u)=p^{(j)}(u)-p^{(k)}(u).
\end{equation}
The regularised analytical source is
\begin{equation}
    \ell_\varepsilon^{(j,k)}(x,u)
    =\gamma_{jk}z_{jk}(u)q_j\!\left(y_j\right)
      W\!\left(y_j,z_{jk};\varepsilon\right).
    \label{eq:appendix-d-analytical-source}
\end{equation}
For the locally frozen front profile, \ref{app:coupling-only} defines the translation mode
\begin{equation}
    v_j(y)=-\partial_y\varphi_j^\star(y),
    \qquad
    N_j=\langle\psi_j,v_j\rangle\ne0,
\end{equation}
and gives the projected motion
\begin{equation}
    \frac{d}{du}p_{\mathrm{coup}}^{(j)}(u)
    =\frac{\langle\psi_j,\ell_\varepsilon^{(j,k)}\rangle}{N_j}
    \simeq-\kappa_{jk}z_{jk}(u).
    \label{eq:appendix-d-projected-law}
\end{equation}
The approximation retains the simple-zero, weak-coupling, local-support and frozen-profile conditions stated in \ref{app:coupling-only}. The raw first moment and its conversion into the ordered rate are kept separate; \ref{app:first-moment-correspondence} treats that relation in detail.

\subsection{Numerical translation mode}

The numerical implementation uses the current receiving-book density on the fixed operational log-price grid. It constructs
\begin{equation}
    v_j^{\Delta x}(x_i,u_n)
    =-\partial_x^{\Delta x}\varphi^{(j)}(x_i,u_n),
    \label{eq:numerical-translation-mode}
\end{equation}
where the derivative is evaluated on the resolved density profile and the outer source values are set to zero. The ordered coupling field acting on book $j$ in response to book $k$ is then
\begin{equation}
    \ell_T^{(j,k)}(x_i,u_n)
    =-\kappa_{jk}z_{jk}(u_n)
      v_j^{\Delta x}(x_i,u_n),
    \qquad \kappa_{jk}\ge0.
    \label{eq:numerical-translation-source}
\end{equation}
If the discrete translation mode is normalised by the same displacement projection, then
\begin{equation}
    \frac{\langle\psi_j,\ell_T^{(j,k)}\rangle}{N_j}
    =-\kappa_{jk}z_{jk}.
    \label{eq:numerical-translation-projection}
\end{equation}
Thus $\kappa_{jk}$ is the ordered numerical boundary-response rate. It is not reconstructed during the simulation from the analytical source amplitude, selector width or raw moment.

\subsection{Common pre-step update}

At operational step $n$, both current density fields and both reaction boundaries are copied before either directed coupling is evaluated. The two ordered fields are therefore
\begin{align}
    \ell_T^{(1,2)}(x_i,u_n)
    &=-\kappa_{12}\bigl(p^{(1)}_n-p^{(2)}_n\bigr)
      v_1^{\Delta x}(x_i,u_n),\\
    \ell_T^{(2,1)}(x_i,u_n)
    &=-\kappa_{21}\bigl(p^{(2)}_n-p^{(1)}_n\bigr)
      v_2^{\Delta x}(x_i,u_n).
    \label{eq:ordered-translation-fields}
\end{align}
They are assembled from the same pre-step state and then included in the simultaneous two-book operational update. This prevents the first updated book from changing the coupling field used for the second book within the same operational step.

For $z_{12}>0$, the first boundary moves downward and the second moves upward. The spread therefore satisfies, to first order,
\begin{equation}
    \frac{d}{du}z_{12}(u)
    =-\bigl(\kappa_{12}+\kappa_{21}\bigr)z_{12}(u),
    \qquad
    \kappa=\kappa_{12}+\kappa_{21}.
    \label{eq:numerical-spread-relaxation}
\end{equation}
The coupling vanishes exactly at zero spread. In the symmetric benchmark, equal ordered rates produce opposite boundary motions and preserve the pair centre to first order.

\subsection{Spatial scales}

The analytical selector regularisation $\varepsilon$, the analytical source scale $\mu_j^{-1/2}$, the resolved numerical reaction-front profile and the lattice spacing $\Delta x$ are distinct objects. In particular, the numerical source in Equation~\eqref{eq:numerical-translation-source} introduces no independent selector width. Its spatial shape is supplied by the current resolved density through $v_j^{\Delta x}$.

Equation~\eqref{eq:appendix-d-analytical-source} remains useful because it gives a bounded source-level construction from which the reduced response can be derived. Equation~\eqref{eq:numerical-translation-source} instead implements that reduced response directly. Hence
\begin{equation}
    \ell_\varepsilon^{(j,k)}
    \not\equiv
    \ell_T^{(j,k)},
    \qquad
    \frac{\langle\psi_j,\ell_\varepsilon^{(j,k)}\rangle}{N_j}
    \simeq
    \frac{\langle\psi_j,\ell_T^{(j,k)}\rangle}{N_j}
    =-\kappa_{jk}z_{jk},
    \label{eq:projected-not-pointwise-correspondence}
\end{equation}
under the local analytical closure and matched ordered response rate.

\subsection{Checks and limitations}

The reduced correspondence requires the correct response under spread reversal, zero coupling at zero spread, recovery of the ordered and total relaxation rates, a resolved interior reaction boundary, and stable behaviour under grid refinement. In the symmetric case the pair centre should also be preserved to first order. These conditions test the reaction-front reduction only within the numerical regime considered.

They do not prove the analytical closure, remove its validity conditions, identify the structural source parameters from $\kappa_{jk}$ alone, or establish pointwise equality between the analytical and numerical fields. Strong coupling, multiple or flat zero crossings, large front deformation, insufficient spatial resolution or material evolution of the front shape on the response timescale remain outside the reduced correspondence.

\subsection{Legacy implementation}

The coupled-book simulation of \citet{BauerDianaGebbie2024} is the numerical antecedent for this work. That implementation used a different source convention, finite-grid front treatment and temporal update. The present implementation instead evolves both books on a uniform operational grid, couples them through the receiving front's translation mode, and applies book-specific previous-refresh clocks only after the operational paths are complete.

\section{First-moment correspondence}
\label{app:first-moment-correspondence}

The analytical first moment and the numerical translation mode enter at different levels. The first moment gives the reduced forcing coefficient under the local reaction-front closure. The numerical implementation starts from the resulting ordered boundary-response rate and applies it directly to the resolved translation mode.

\subsection{Selected first moment}

Write
\begin{equation}
    y=x-p^{(j)},
    \qquad
    z=z_{jk}=p^{(j)}-p^{(k)},
\end{equation}
and let the analytical source-shape kernel be
\begin{equation}
    q_j(y)=-\lambda_j\mu_j y e^{-\mu_jy^2}.
    \label{eq:first-moment-source-kernel}
\end{equation}
The regularised and hard-side continuum sources are
\begin{align}
    \ell_\varepsilon^{(j,k)}(y,z)
    &=\gamma_{jk}zq_j(y)W(y,z;\varepsilon),\\
    \ell_0^{(j,k)}(y,z)
    &=\gamma_{jk}zq_j(y)\mathbf 1_{\{yz>0\}},
\end{align}
where
\begin{equation}
    W(y,z;\varepsilon)
    =\frac12\left[1+\tanh\!\left(\frac{yz}{\varepsilon}\right)\right].
    \label{eq:first-moment-selector}
\end{equation}
These fields are not pointwise identical for finite $\varepsilon$. Their selected raw first moment is
\begin{equation}
    \mathcal M_j[\ell](z)
    =\int_{\mathbb R}y\ell(y,z)\,dy.
\end{equation}
Assume that $q_j$ is odd and integrable with a finite first moment, and that the local integration domain is symmetric. Then $yq_j(y)$ is even, while
\begin{equation}
    yq_j(y)\tanh\!\left(\frac{yz}{\varepsilon}\right)
\end{equation}
is odd in $y$. Hence, for every finite $\varepsilon>0$,
\begin{align}
    \int_{\mathbb R}yq_j(y)W(y,z;\varepsilon)\,dy
    &=\frac12\int_{\mathbb R}yq_j(y)\,dy\\
    &=\int_0^\infty yq_j(y)\,dy
    =:M_j^+.
\end{align}
The hard-side source gives the same selected moment for either sign of $z$. Therefore
\begin{equation}
    \mathcal M_j[\ell_\varepsilon^{(j,k)}](z)
    =\mathcal M_j[\ell_0^{(j,k)}](z)
    =\gamma_{jk}M_j^+z.
    \label{eq:exact-raw-moment}
\end{equation}
This identity is exact under the parity and integrability assumptions and requires no $\varepsilon\to0$ limit.

For Equation~\eqref{eq:first-moment-source-kernel},
\begin{align}
    M_j^+
    &=-\lambda_j\mu_j\int_0^\infty y^2e^{-\mu_jy^2}\,dy\\
    &=-\frac{\lambda_j\sqrt\pi}{4\sqrt{\mu_j}}<0,
    \label{eq:gaussian-positive-half-moment}
\end{align}
where
\begin{equation}
    \int_0^\infty y^2e^{-\mu y^2}\,dy
    =\frac{\sqrt\pi}{4\mu^{3/2}}.
\end{equation}

\subsection{Boundary-response rate}

The raw moment does not by itself give a reaction-boundary velocity. \ref{app:coupling-only} introduces the adjoint response mode and the normalised projection
\begin{equation}
    \frac{d}{du}p_{\mathrm{coup}}^{(j)}
    =\frac{\langle\psi_j,\ell_\varepsilon^{(j,k)}\rangle}{N_j}.
\end{equation}
Under the local first-moment approximation
\begin{equation}
    \frac{\psi_j(y)}{N_j}\simeq A_jy,
\end{equation}
Equations~\eqref{eq:exact-raw-moment} and \eqref{eq:first-moment-front-closure} give
\begin{align}
    \frac{d}{du}p_{\mathrm{coup}}^{(j)}
    &\simeq A_j\gamma_{jk}M_j^+z_{jk}
    =-\kappa_{jk}z_{jk},\\
    \kappa_{jk}
    &:=-A_j\gamma_{jk}M_j^+>0.
    \label{eq:first-moment-ordered-rate}
\end{align}
For the signed-density convention used here, $A_j>0$, $\gamma_{jk}>0$ and $M_j^+<0$, giving a positive mean-reverting ordered rate. The factor $A_j$ carries the normalisation, orientation and dimensions of the front projection. If $\varphi^{(j)}$ is signed order volume per unit log-price, then $A_j$ has units of inverse order volume and the projected velocity has units of log-price per unit operational time.

The structural map is therefore
\begin{equation}
    (A_j,\gamma_{jk},\lambda_j,\mu_j)
    \longmapsto
    \kappa_{jk}
    =A_j\gamma_{jk}
      \frac{\lambda_j\sqrt\pi}{4\sqrt{\mu_j}}.
    \label{eq:structural-to-response-rate}
\end{equation}
This map is many-to-one. A measured or declared $\kappa_{jk}$ does not separately identify the source amplitude, source scale, coupling strength and front-response normalisation.

\subsection{Numerical response}

The active numerical implementation does not evaluate Equation~\eqref{eq:structural-to-response-rate} during the path simulation. It declares $\kappa_{jk}$ as the ordered response rate and applies
\begin{equation}
    \ell_T^{(j,k)}(x_i,u_n)
    =-\kappa_{jk}z_{jk}(u_n)
      v_j^{\Delta x}(x_i,u_n),
    \qquad
    v_j^{\Delta x}=-\partial_x^{\Delta x}\varphi^{(j)}.
    \label{eq:first-moment-numerical-translation-source}
\end{equation}
As shown in \ref{app:numerical-reaction-front-correspondence}, projection of this field gives the ordered law $\dot p_j=-\kappa_{jk}z_{jk}$ directly. Thus
\begin{equation}
    (A_j,\gamma_{jk},\lambda_j,\mu_j)
    \longmapsto
    \kappa_{jk}.
    \label{eq:first-moment-theory-code-map}
\end{equation}
The analytical closure gives $\kappa_{jk}$, which is then prescribed in the numerical coupling and recovered from the boundary motion. The two source fields need not be pointwise equal, and $\kappa_{jk}$ does not separately identify the structural parameters.

\subsection{Scales and limits}

Since $y$ and $z$ both have units of log-price,
\begin{equation}
    [\varepsilon]=[\mathrm{log\text{-}price}]^2.
\end{equation}
For fixed $z\ne0$, the selector transition width in the local coordinate is
\begin{equation}
    w_\varepsilon(z)\sim\frac{\varepsilon}{|z|}.
\end{equation}
A reference-scale parameterisation is
\begin{equation}
    \varepsilon=|z|_{\mathrm{ref}}w_{\mathrm{ref}},
    \qquad
    \Delta x\ll w_{\mathrm{ref}}\ll L_{q,j},
    \qquad
    L_{q,j}\sim\mu_j^{-1/2}.
\end{equation}
The condition $\Delta x\ll\sqrt\varepsilon\ll L_{q,j}$ follows only under the symmetric convention $|z|_{\mathrm{ref}}=w_{\mathrm{ref}}=\sqrt\varepsilon$.

The pointwise limits $z\to0$ and $\varepsilon\to0$ do not commute: fixed $\varepsilon$ gives $W\to1/2$ as $z\to0$, whereas the hard-side limit at fixed nonzero $z$ gives $W\to\mathbf 1_{\{yz>0\}}$. This does not change Equation~\eqref{eq:exact-raw-moment} under the stated parity and integrability assumptions. It does prevent the regularisation from being interpreted as a pointwise rewrite of the numerical translation-mode field.

The numerical implementation introduces no $\varepsilon$ or independent selector width. Its front shape and width are supplied by the resolved density. The analytical selector scale, analytical source scale, resolved numerical front and lattice spacing must therefore remain distinct.

\end{document}